# Photonic devices fabricated from (111) oriented single crystal diamond

**Blake Regan[1], Sejeong Kim[1], Anh Tu Huy Ly[1], Aleksandra Trycz[1], Kerem Bray[1], Kumaravelu Ganesan[2], Milos Toth[1], and Igor Aharonovich[1]**

*1. School of Mathematical and Physical Sciences, University of Technology Sydney, Ultimo, New South Wales 2007, Australia*
*2. School of Physics, University of Melbourne, Vic 3010, Australia*

**Abstract**
*Diamond is a material of choice in the pursuit of integrated quantum photonic technologies. So far, the majority of photonic devices fabricated from diamond, are made from (100)-oriented crystals. In this work, we demonstrate a methodology for the fabrication of optically-active membranes from (111)-oriented diamond. We use a liftoff technique to generate membranes, followed by chemical vapour deposition of diamond in the presence of silicon to generate homogenous silicon vacancy colour centers with emission properties that are superior to those in (100)-oriented diamond. We further use the diamond membranes to fabricate high quality microring resonators with quality factors exceeding ~ 3000. Supported by finite difference time domain calculations, we discuss the advantages of (111) oriented structures as building blocks for quantum nanophotonic devices.*

Diamond is an attractive platform for studies of light-matter interaction at the nanoscale[1-7]. In the past decade, defects in diamond, also known as color centers, have emerged as attractive candidates for scalable solid state quantum photonic architectures[1, 6]. While earlier works were focused predominantly on nitrogen vacancy (NV) centres[8, 9], recent effort is devoted to group IV defects[10-15], (e.g. the silicon vacancy (SiV)) due to their narrowband emission and a better resilience to electromagnetic fluctuations.

Integration of these color centres with photonic resonators is an important challenge for several reasons. First, it enables enhancement of the photon emission flux from the single color centres (for instance by using diamond pillars[16]). Second, it provides the means to interconnect potential quantum nodes in a large network, where the emitters are coupled to individual cavities and interconnected with a waveguide[17, 18]. Last, if the photonic resonator is carefully designed with coupled directional emission, one can achieve high cavity cooperativity and realise advanced quantum phenomena such as single photon switch with a solid state system[19]. However, despite the remarkable progress in nanofabrication of diamond cavities[8, 20], all photonic resonators to date were fabricated from (100)-oriented diamond, primarily because this is the most common orientation provided by commercial suppliers and the difficulty in engineering and polishing (111) oriented crystals[21]. This, however, limits the potential coupling strength of many color centers to cavities since most centers studied to date have dipoles oriented along the <111> direction, and the dipole overlap with the cavity field is therefore not optimal in cavities fabricated from (100)-oriented diamond. Conversely, in (111)-oriented diamond, the overlap is optimal, and superior Purcell enhancement is expected.

In the current work, we fabricated large area (111) diamond membranes that exhibit bright SiV luminescence. Consequently, these membranes were utilized to engineer high quality diamond microring resonators with quality factors of ~ 3000. Our work launches a new approach for the fabrication of diamond devices with different crystallographic orientations.

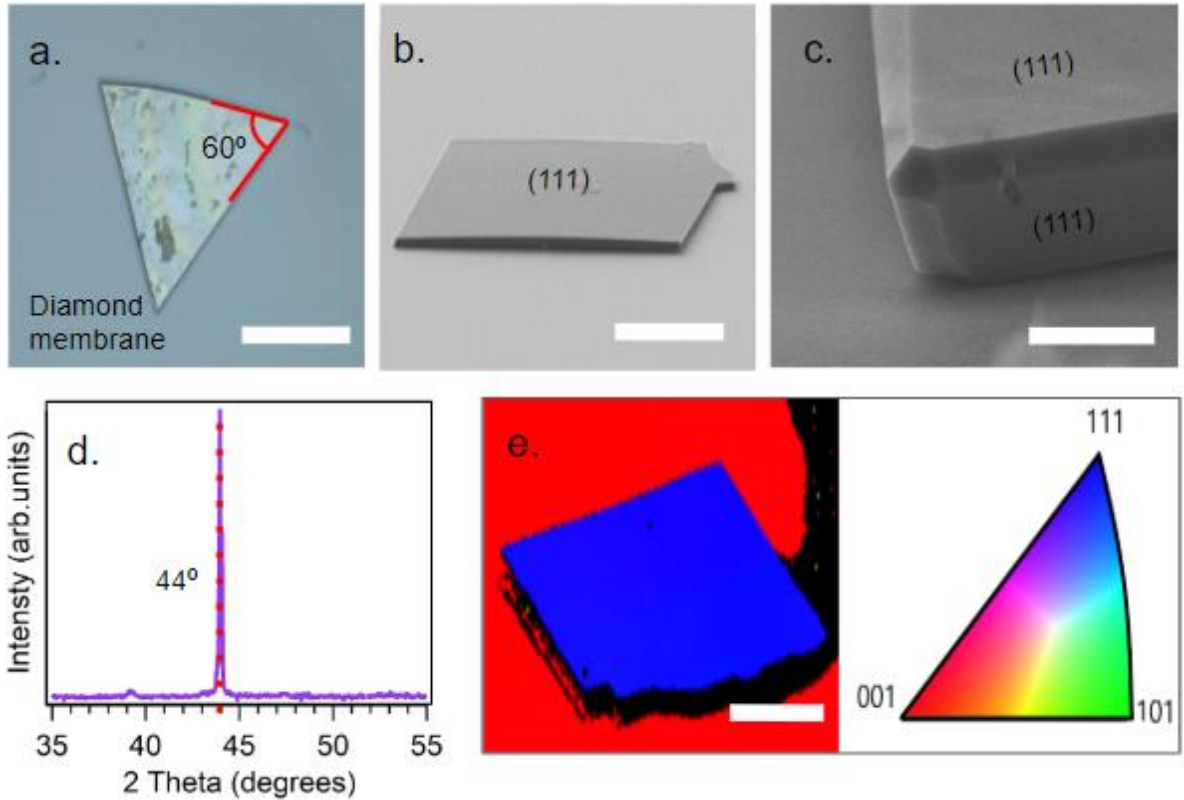

**Figure 1**. **Single crystal (111) diamond membranes.** (a) Optical image a (111) membrane showing triangular cleaving angle, marked at ~ 60°. (b) SEM image in tilted view showing a (111) membrane. (c) Magnified view of a membrane edge after CVD overgrowth. The scale bar corresponds to 1μm. (d) X-ray diffraction (XRD) spectrum showing a peak at 44°, characteristic of a (111)-orientated crystal. (e) EBSD map of a (111)-oriented, overgrown diamond membrane (blue). The red background is (001)-oriented silicon, and the dark regions are amorphous PMMA. The image was taken at an electron beam incidence angle of 70°. The scale bars in a, b, e correspond to 100 μm.

The fabrication process begins with (111)-oriented single crystal diamond (obtained from New Diamond Technology). Single-crystal membranes were fabricated by helium ion implantation, followed by an electrochemical liftoff process[22, 23]. Figure 1a shows an optical image of a diamond membrane after liftoff. In many cases, the shape of the membrane is defined by the (111) orientation in the face centered cubic structure, characterised by a cleavage angle of ~60°. Figure 1b shows a scanning electron microscope (SEM) image of a different diamond membrane.

The membranes are then transferred onto a silicon substrate and overgrown using a microwave plasma chemical vapor deposition (MPCVD) system to introduce SiV centres[24]. The MPCVD growth conditions are as follows: a hydrogen/methane ratio of 100:1 at 60 Torr, a microwave power of 900W, and a growth time of 10 minutes. Figure 1c shows an overgrown diamond membrane, with a smooth, single crystal surface.

To confirm the (111) membrane orientation, X-ray diffraction (XRD) and electron backscattered diffraction (EBSD) analyses were performed. Figure 1d shows an XRD spectrum of the original bulk single crystal diamond with a pronounced peak at $2\theta = 44°$ degrees, confirming the {111 orientation of the crystal[21]. Figure 1e shows an EBSD map recorded from the overgrown membrane, confirming further that it is single crystal (111) oriented diamond through analysis of kikuchi pattern information derived from back scattering of electrons in SEM.

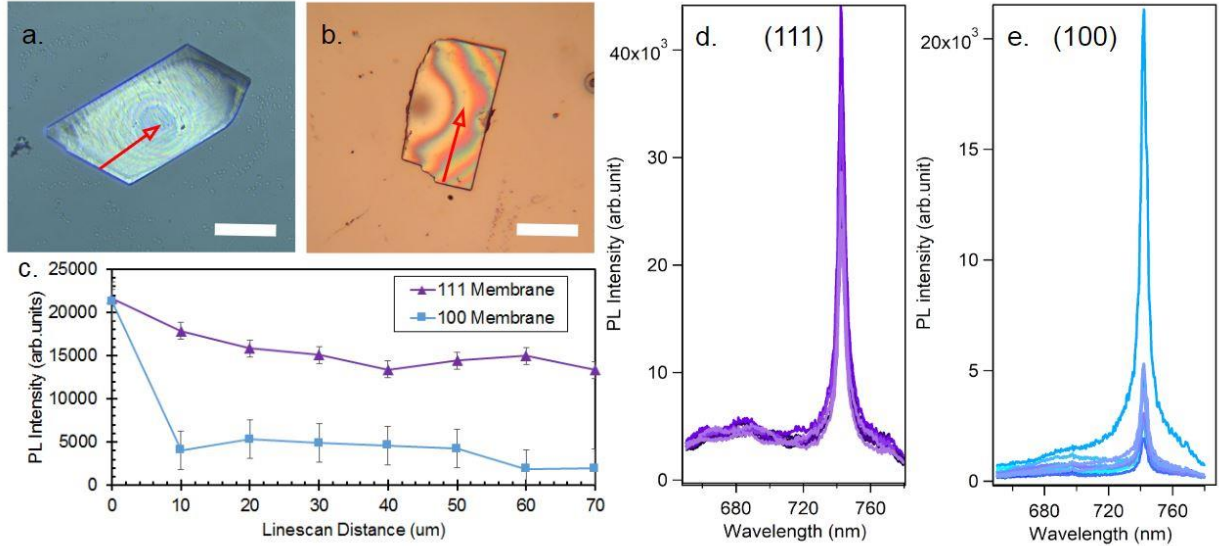

**Figure 2. Homogeneous distribution of SiV centres in a 111-oriented diamond membrane.** Optical images showing (a) a (111) diamond membrane and (b) (100) diamond membrane. Scale bars correspond to 50 μm. Red arrows indicate line scan direction and length used for (d) and (e). (c) PL intensity versus distance for (111) and (100)-oriented membranes. Photoluminescence spectra collected along the line scans from the (111)-oriented membrane (d) and (100)-oriented membrane (e).

A comparison of (111) and (100) oriented membranes was performed to demonstrate the advantages of the former. A (111)-oriented diamond membrane and a similarly sized (100)-membrane were situated on a silicon substrate ~100 μm apart. Both membranes were cleaned in piranha solution prior to fabrication. The membranes were overgrown simultaneously using the MPCVD conditions detailed above in order to incorporate SiV colour centers (see figure 2a, 2b). Each membrane was subsequently flipped and thinned down to 300 nm using ICP-RIE, and their thickness was confirmed with high angle SEM analysis. The thinned membranes were subsequently examined by photoluminescence (PL) spectroscopy. While the intensity at the edge of each membrane was comparable, line scans (figure 2c) across the membranes show a clear difference in PL homogeneity. The average PL intensity from the (111)-oriented membrane (figure 2a) is more intense, and the line profile is more uniform. The (100) membrane, on the other hand, (figure 2b) showed a significantly higher intensity reduction moving inward from the edges of the membrane. PL spectra from both membranes are shown in figure 2d, e, showing a prominent SiV emission with a zero phonon line (ZPL) at 738 nm. Given the similar thicknesses of the membranes after the thinning step, the difference in PL intensity is attributed to a better incorporation of the SiV into the growing (111) oriented diamond membrane. Similar results were obtained previously for the NV centers in diamond that showed to preferentially incorporate into (111) grown diamonds[25-28]. The edges of both membranes have similar PL intensities since a (100)-oriented membrane has a (111) facet at the edge due to the three dimensional single crystal diamond growth. Both membranes do experience a dropoff in intensity toward the center of the membrane, this can be attributed to scattering enhancement at the edges and losses into bulk material.

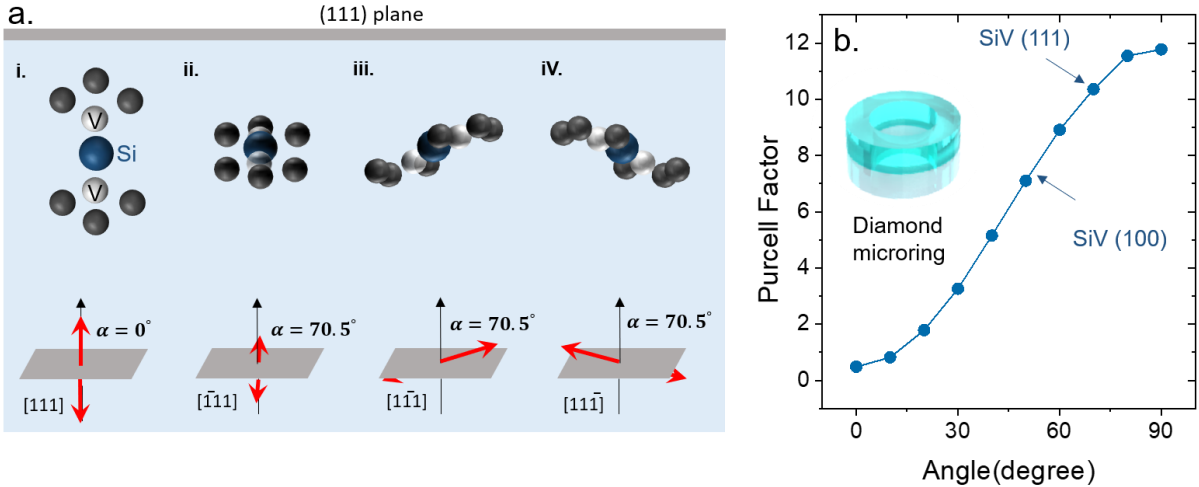

**Figure 3**. **Dipole orientation in (111)-oriented diamond and Purcell factors.** (a) Simplified schematics of the four possible orientations of SiV centers in (111)-oriented diamond and the corresponding dipole orientations (red arrows). The angle α between the <111> direction and the dipole is shown in each case. (b) Calculated Purcell factor as a function of a dipole angle with respect to the top surface of the ring.

To provide more insight into the coupling efficiency between the emitters in diamond membrane and the cavity modes, we perform a finite-difference time domain (FDTD) simulation. We focus here on the group IV defects, and predominantly the SiV defect. Its structure is a split vacancy with a silicon in an interstitial location along the <111> symmetry axis, resulting in a $D_3d$ point group symmetry. Figure 3a shows the schematic drawing of the four (i,ii, iii and iv) potential orientation of SiV defect in a (111) oriented diamond[29, 30]. These orientations are further simplified with red arrows indicating dipole orientations and α indicating angles between the dipole and the <111> vector. One SiV dipole orientation is perpendicular to the surface, while the rest of the dipoles make equal angle of 70.5 degree with the top diamond surface. The same SiV dipole in (100) membrane would form an angle of 54.7 degrees. Most of the photonic cavities are designed for transverse electric (TE)-like modes, meaning dipoles with larger angle α would have a better electric field overlap. Consequently, Purcell factor with respect to the dipole angle α is calculated as shown in Figure 3b for a diamond microring. The simulation result shows that more in-plain components dipole has the higher Purcell factor is expected. Note that we conducted simulation for SiV centre in this case, but this would be true also for other group IV defects in diamond.

Given the advantages of the (111) membranes, ring resonators have been engineered from this material. Electron beam lithography was employed to pattern a hydrogen silsesquioxane (HSQ) layer spuncoat onto the diamond membrane, forming an $SiO_2$ hard mask array of microring resonators with external diameter of ~ 400 nm and a radius of 5 μm. The pattern is transferred into the diamond membrane through a second ICP-RIE step with Oxygen at 10 mTorr. Finally, the mask is removed with another RIE step of SF6 at 40 mTorr to directly etch the $SiO_2$ mask and undercut into the $SiO_2$ substrate. Figure 4a shows the SEM of the patterned microrings, while figure 4b shows a high resolution image of an individual microring.

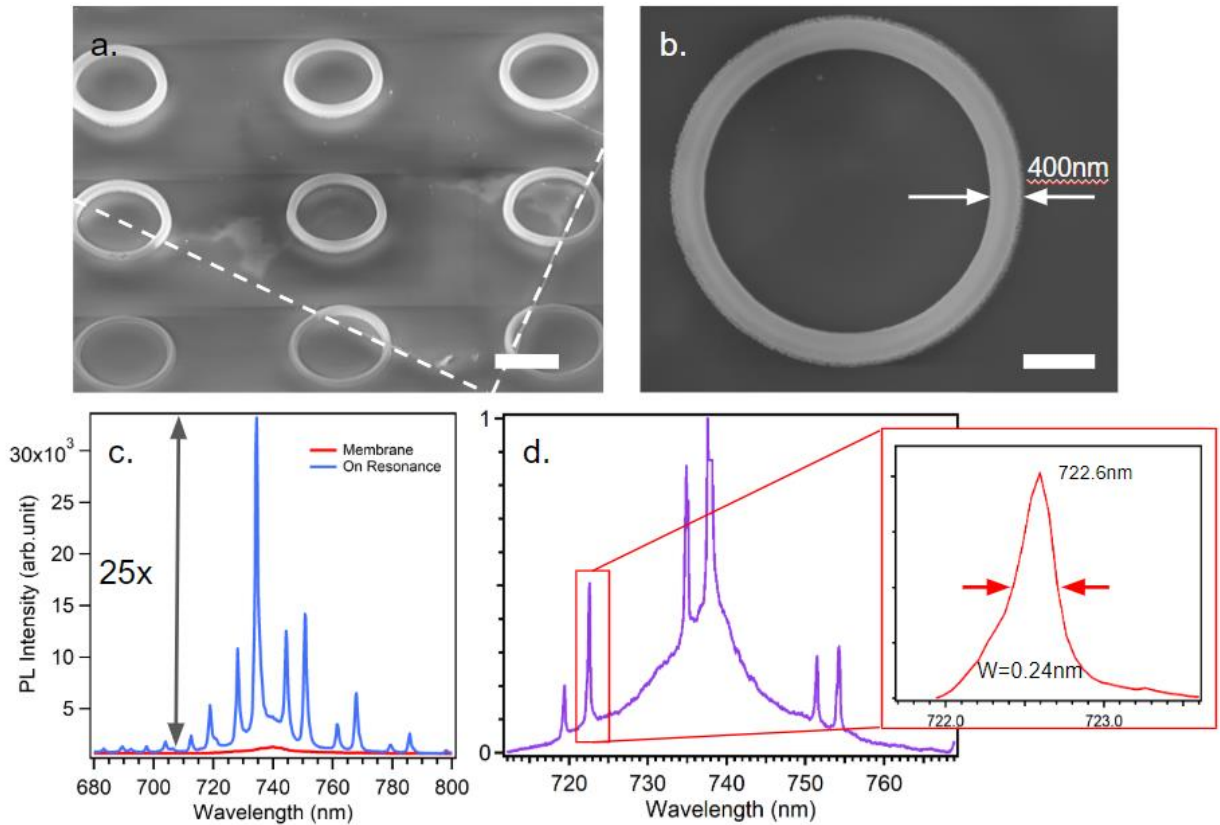

**Figure 4. Diamond ring resonators fabricated from (111) diamond membrane containing SiV colour centers.** (a) shows a wide area at 45 degree inclination depicting the array of microrings. The doted line is the membrane boundary, scale bar corresponds to 3 μm (b) Shows a top view of an individual microring, scale bar correlates to 1 μm. (c) PL spectrum recorded from the diamond microring cavity under 532 nm excitation at room temperature. WGMs are clearly visible. The ZPL of the SiV center is observed at 738 nm. Comparison of recorded emission from SiV color center recorded from the membrane and from a micro-ring cavity, showing an enhancement of ~ 25 times. (d) fine grating Spectral analysis of whispering gallery resonators, individual marked mode with FWHM of ~ 0.24 nm, Q factor ~3000.

The diamond cavities were then probed using a green laser in a standard confocal microscope. Figure 4c, shows the whispering gallery modes (WGMs) from the resonator (manifested by the periodic peaks in the spectrum). With the designated microring size, the free spectral range should be ~ 14 nm. The spectrum of the ring resonators is augmented by the modes of the optical cavity surrounding the SiV peak at ~ 738nm. Figure 4c further shows an emission enhancement form the fabricated microring cavities as compared to the membrane. An enhancement of ~25 times is observed after measuring a number of devices.

A high resolution spectrum centered at the SiV region is shown in figure 4d. Quality factors as high as ~3000 were observed from these devices (measured as $\lambda/\lambda\Delta$). The losses occur due to an increased roughness from the imperfect HSQ mask removal and from the leakage to the underlying SiO2 substrate.

The enhancement of brightness and high quality factor are desirable and advantageous properties for the application of colourcenters to practical nanophotonic devices.

In conclusion, we have described a robust method to engineer high-quality single crystal diamond membranes from (111) oriented diamond. The improved brightness in (111)-oriented membranes is favorable for sensing applications - particularly for NV based magnetometry, and for photonic devices with group IV color centres. Furthermore, we have demonstrated the use of this platform for the fabrication of ring resonators with quality factors up to ~3000, supported by a theoretical model that confirms a higher Purcell enhancement for the (111) oriented devices. Similar techniques can be explored to achieve doping of (111) diamond membranes with other group IV emitters[15], or other emerging color centres in diamond. Our results will accelerate the integration of diamond with scalable photonic devices to achieve on-chip quantum nanophotonic circuitry.

**Acknowledgements**

We thank Dr Carlo Bradac for useful discussions and Dr Mark Lockrey for assistance with the EBSD data. Financial support from the Australian Research Council (via DP180100077, DP190101058), the Asian Office of Aerospace Research and Development (grant FA2386-17-1-4064), and the Office of Naval Research Global (grant N62909-18-1-2025).